\documentclass{article}

\usepackage{amsthm}
\usepackage[english]{babel}

\theoremstyle{plain}
\theoremstyle{definition}
\newtheorem{definition}{Definition}
\newtheorem{example}{Example}
\newtheorem{lemma}{Lemma}
\newtheorem{corollary}{Corollary}
\newtheorem{theorem}{Theorem}

\theoremstyle{remark}

\newcommand{\cc}{\mbox{C}}

\newcommand{\p}{\mbox{P}}

\newcommand{\np}{\mbox{NP}}

\newcommand{\N}{\mathbb{N}}

\newcommand{\qfbapa}{\textsf{QFBAPA}}

\newcommand{\parikh}{\textsf{Parikh}}

\usepackage{amssymb}

\usepackage[utf8]{inputenc}
\usepackage{mathtools}
\usepackage{xparse}

\DeclarePairedDelimiterX{\set}[1]{\{}{\}}{\setargs{#1}}
\NewDocumentCommand{\setargs}{>{\SplitArgument{1}{;}}m}
{\setargsaux#1}
\NewDocumentCommand{\setargsaux}{mm}
{\IfNoValueTF{#2}{#1} {#1\nonscript\:\delimsize\vert\allowbreak\nonscript\:\mathopen{}#2}}
\def\Set{\set*}

\DeclarePairedDelimiterX{\mset}[1]{\llparenthesis}{\rrparenthesis}{\msetargs{#1}}
\NewDocumentCommand{\msetargs}{>{\SplitArgument{1}{;}}m}
{\msetargsaux#1}
\NewDocumentCommand{\msetargsaux}{mm}
{\IfNoValueTF{#2}{#1} {#1\nonscript\:\delimsize\vert\allowbreak\nonscript\:\mathopen{}#2}}

\usepackage[affil-it]{authblk}

\usepackage{graphicx}

\usepackage{cite}

\usepackage[inline]{enumitem}

\usepackage{ amsmath }

\usepackage{ amsfonts }

\usepackage{ stmaryrd }

\usepackage{graphicx}
\usepackage{float}
\usepackage{tikz}
\usetikzlibrary{positioning,automata} 

\tikzset{%
  zeroarrow/.style = {-stealth,dashed},
  onearrow/.style = {-stealth,solid},
  c/.style = {circle,draw,solid,minimum width=2em,
        minimum height=2em},
  r/.style = {rectangle,draw,solid,minimum width=2em,
        minimum height=2em}
}

\usepackage{booktabs}

\usepackage[hidelinks]{hyperref}

\usepackage[strings]{underscore}

\usepackage{braket}

\begin{document}

\title{The Complexity of Satisfiability Checking for Symbolic Finite Automata}


\author{Rodrigo Raya}


\affil{School of Computer and Communication Sciences, EPFL, Switzerland }


\maketitle              

\begin{abstract}
We study the satisfiability problem of symbolic finite automata and decompose it into the satisfiability problem of the theory of the input characters and the monadic second-order theory of the indices of accepted words. We use our decomposition to obtain tight computational complexity bounds on the decision problem for this automata class and an extension that considers linear arithmetic constraints on the underlying effective Boolean algebra. 

\end{abstract}

\section{Introduction}
\label{section:intro}
Symbolic finite automata (SFAs) are an extension of finite automata that allow transitions to be  labelled with monadic predicates over some universe rather than symbols from a finite alphabet. They were first mentioned in \cite{watson_implementing_1996}, but they attracted renewed interest starting in \cite{veanes_rex_2010}. SFAs have been used in a variety of applications including the analysis of regular expressions \cite{veanes_rex_2010, dantoni_minimization_2014}, string encoders,  sanitizers \cite{hooimeijer_fast_2011, dantoni_extended_2015, hu_automatic_2017}, functional programs \cite{dantoni_fast_2015}, code generation, parallelization \cite{saarikivi_fusing_2017} and symbolic matching \cite{saarikivi_symbolic_2019}. 

A series of theoretical investigations has been carried out on this automata model, including \cite{dantoni_minimization_2014, tamm_theoretical_2018, argyros_learnability_2018}. In particular, the authors of \cite{veanes_monadic_2017} observed that such an automata model had been studied previously by Bès in  \cite{bes_application_2008}. In his paper, Bès introduced a class of  multi-tape synchronous finite automata whose transitions are labelled by first-order formulas. He then proved various properties of the languages accepted by such automata including closure under Boolean, rational, and the projection operations, logical characterizations in terms of MSO logic and the Eilenberg-Elgot-Shepherdson formalism as well as decidability properties. Remarkably \cite{veanes_monadic_2017}, the paper showed that recognizability for such automata coincides with definability for certain generalized weak powers, first-studied by Feferman and Vaught in \cite{feferman_first_1959}.

The techniques of Feferman and Vaught allow decomposing the decision problem for the first-order theory of a product of structures, $Th(\prod_i \mathcal{M}_i)$ into the first-order theory of the structures $\mathcal{M}_i$, $Th(\mathcal{M}_i)$, and the monadic second-order theory of the index set $I$, $Th^{mon}(\langle I, \ldots \rangle)$, where the structure $\langle I, \ldots \rangle$ may contain further relations such as a finiteness predicate, a cardinality operator, etc. If the theory of the components $Th(\mathcal{M}_i)$ is decidable, then the decision problem reduces to that of the theory $Th^{mon}(\langle I, \ldots \rangle)$. To analyse these structures, Feferman and Vaught extend results that go back to Skolem \cite{skolem_untersuchungen_1919}. Technically, the decomposition is expressed in terms of so-called reduction sequences. 

\sloppy{
It is known \cite{dawar_model_2007} that many model-theoretic constructions incur in non-elementary blow-ups in the formula size. This includes the case of the size of the Feferman-Vaught reduction sequences in the case of disjoint unions. Perhaps for this reason, no computational complexity results have been obtained for the theory of symbolic automata and related models. Instead, the results in the literature \cite{dantoni_power_2017, fisman_complexity_2021, dantoni_automata_2021} refer to the decidability of the satisfiability problem of the monadic predicates or provide asymptotic run-times rather than a refined computational complexity classification. 
}

As a \textbf{main contribution}, we show how to reduce the satisfiability problem for finite symbolic automata to the satisfiability problem of the existential first-order theory of the theory of the elements and the existential monadic second-order theory of the indices. This decomposition allows us to derive tight complexity bounds for the decision problem of the automaton in the precise sense of Corollary~\ref{cor:complexity}. We then study an extension of the formalism of symbolic finite automata which also imposes linear arithmetic constraints on the cardinalities of the Venn regions of the underlying effective Boolean algebra. In particular, this extension allows expressing the number of
occurrences of a particular kind of letter in a word. We show in Corollary~\ref{cor:complexitycard} that the computational complexity of the corresponding satisfiability problem is the same as the one for the simpler model without cardinalities. Similar extensions for related models of automata are considered in the literature \cite{figueira_reasoning_2022}.

\textbf{Organisation of the paper.} Section~\ref{section:symbolic} introduces symbolic finite automata. Section~\ref{section:feferman} gives the Feferman-Vaught decomposition of symbolic finite automata in terms of the theory of the elements and the theory of the indices. Section~\ref{section:decision} describes the decision procedure with which, in Section~\ref{section:with}, after presenting the quantifier-free theory of Boolean algebra with Presburger arithmetic, we obtain the tight complexity bounds announced. Section~\ref{section:cardinalities} describes the extension of symbolic finite automata that uses linear arithmetic constraints over the cardinalities of the automaton's underlying effective Boolean algebra and proves the corresponding upper bounds for the associated satisfiability problem. Section~\ref{section:conclusion} concludes the paper.

\section{Symbolic Finite Automata (SFA)}
\label{section:symbolic}
Symbolic automata are run over Boolean algebras of interpreted sets. The family of monadic predicates used for these interpretations needs to be closed under Boolean operations and contain formulae denoting the empty set and the universe. Furthermore, in the original formulation, checking non-emptiness of these interpreted sets needs to be decidable. In Section~\ref{section:decision}, we will refine this assumption with a complexity-theoretic bound.

\begin{definition}[\cite{dantoni_automata_2021}]
An effective Boolean algebra $\mathcal{A}$ is a tuple 
\[
\left(\mathfrak{D}, \Psi, \llbracket \cdot \rrbracket, \perp, \top, \vee, \wedge, \neg\right)
\]
where $\mathfrak{D}$ is a set of domain elements, $\Psi$ is a set of unary predicates over $\mathcal{D}$ that are closed under the Boolean connectives, with $\perp, \top \in \Psi$ and $\llbracket \cdot \rrbracket: \Psi \rightarrow 2^{\mathcal{D}}$ is a function such that \begin{enumerate*}
\item $\llbracket \perp \rrbracket=\emptyset$, 
\item $\llbracket \top \rrbracket=\mathfrak{D}$, and 
\item For all $\psi, \psi_1, \psi_2 \in \Psi$, we have that
\begin{enumerate*}
\item $\llbracket \psi_1 \vee \psi_2 \rrbracket=\llbracket \psi_1 \rrbracket \cup \llbracket \psi_2 \rrbracket$
\item $\llbracket \psi_1 \wedge \psi_2 \rrbracket=\llbracket \psi_1 \rrbracket \cap \llbracket \psi_2 \rrbracket$
\item $\llbracket \lnot \psi \rrbracket=\mathfrak{D} \backslash \llbracket \psi \rrbracket$. 
\end{enumerate*}
\item Checking $\llbracket \psi \rrbracket \neq \emptyset$ is decidable. 
\end{enumerate*} 
A predicate $\psi \in \Psi$ is atomic if it is not a Boolean combination of predicates in $\Psi$. 
\end{definition}

Our initial motivation was to generalise the complexity results obtained for array theories 
in \cite{alberti_cardinality_2017, raya_vmcai_2022}. The notion of SMT algebra \cite[Example~2.3]{dantoni_automata_2021} precisely corresponds to the language introduced in \cite[Definition~5]{raya_vmcai_2022} without cardinality constraints. We take this as a first example of effective Boolean algebra. 

\begin{example} \label{example:smt}
The SMT algebra for a type $\tau$ is the tuple $(\mathcal{D}, \Psi, \llbracket \cdot \rrbracket, \bot, \top, \lor, \land, \lnot)$ where $\mathcal{D}$ is the domain of $\tau$, $\Psi$ is the set of all quantifier-free formulas with one fixed free variable of type $\tau$, $\llbracket \cdot \rrbracket$ maps each monadic predicate to the set of its satisfying assignments, $\bot$ denotes the empty set, $\top$ denotes the universe $\mathcal{D}$ and $\lor, \land, \lnot$ denote the Boolean algebra operations of union, intersection, and complement respectively. 
\end{example}

This example should be contrasted with other representations of the predicates that take into account implementation details. An example of the latter is the $k$-bit bitvector effective Boolean algebra described in \cite{konstantinidis_applications_2013}.

\begin{example} \label{example:bit}
The powerset algebra $2^{bv(k)}$ is the tuple $(D,\Psi, \llbracket \cdot \rrbracket, \bot, \top, \lor, \land, \lnot)$ where $\mathcal{D}$ is the set $bv(k)$ of all non-negative integers less than $2^k$ or equivalently, all $k$-bit bit-vectors for some $k > 0$, $\Psi$ is the set of BDDs of depth $k$, $\llbracket \cdot \rrbracket$ maps each BDD $\beta$ to the set of all integers $n$ such that the binary representation of $n$ is a solution of $\beta$, $\bot$ denotes the BDD representing the empty set, $\top$ denotes the BDD representing the universal set and $\lor, \land, \lnot$ denote the Boolean algebra operation of union, intersection, and complement as they are implemented for BDDs.
\end{example}

We now introduce the automata model we will investigate in the paper. 

\begin{definition}[\cite{dantoni_automata_2021}] \label{def:sfa}
A symbolic finite automaton (s-FA) is a tuple 
\[
M=\left(\mathcal{A}, Q, q_{0}, F, \Delta\right)
\]
where \begin{enumerate*}
\item $\mathcal{A}$ is an effective Boolean algebra.
\item $Q$ is a finite set of states. 
\item $q_{0} \in Q$ is the initial state.
\item $F \subseteq Q$ is the set of final states. 
\item $\Delta \subseteq Q \times \Psi_{\mathcal{A}} \times Q$ is a finite set of transitions.
\end{enumerate*}

A symbolic transition $\rho=\left(q_{1}, \psi, q_{2}\right) \in \Delta$, also denoted $q_{1} \stackrel{\psi}{\rightarrow} q_{2}$, has source state $q_{1}$, target state $q_{2}$, and guard $\psi$. For $d \in \mathfrak{D}$, the concrete transition $q_{1} \stackrel{d}{\rightarrow} q_{2}$ denotes that $q_{1} \stackrel{\psi}{\rightarrow} q_{2}$ and $d \in \llbracket \psi \rrbracket$ for some $\psi$.

A string $w=d_{1} d_{2} \ldots d_{k}$ is accepted at state $q$ if and only if for $1 \leq i \leq k$, there exist transitions $q_{i-1} \stackrel{d_{i}}{\rightarrow} q_{i}$ such that $q_{0}=q$ and $q_{k} \in F$. The set of strings accepted at $q$ is denoted by $\mathcal{L}_{q}(M)$ and the language of $M$ is $\mathcal{L}(M)=\mathcal{L}_{q_0}(M)$.
\end{definition}

We now give examples of automata running over the algebras of Examples~\ref{example:smt} and \ref{example:bit}. We use the traditional graphical representation used in automata theory textbooks \cite{hopcroft_introduction_2006}. 

\begin{example}[\cite{dantoni_automata_2021}] 
\label{ex:smt-automaton}
We consider the language of linear arithmetic over the integers. We set two formulae $\psi_{> 0}(x) \equiv x > 0$ satisfied by all positive integers and $\psi_{\text{odd}}(x) \equiv x \mod 2 = 1$ satisfied by all odd integers. The following symbolic finite automaton accepts all strings of even length consisting only of positive odd numbers.

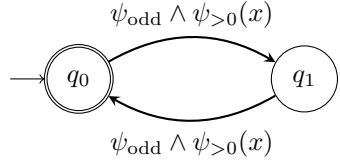
\centering
\begin{tikzpicture} [
    node distance = 3cm, 
    on grid, 
    auto,
    every loop/.style={stealth-}]
 
\node (q0) [state, 
    initial, 
    accepting, 
    initial text = {}] {$q_0$};
 
\node (q1) [state,
    right = of q0] {$q_1$};
 
\path [-stealth, thick]
    (q0) edge[bend left] node {$\psi_{\text{odd}} \land \psi_{> 0}(x)$}   (q1)
    (q1) edge[bend left] node {$\psi_{\text{odd}} \land \psi_{> 0}(x)$}   (q0);
\end{tikzpicture}
\end{example}

\newsavebox{\mydiagram}
\sbox{\mydiagram}{
    \begin{tikzpicture}[node distance=.3cm and .3cm]\footnotesize
    \node (b6) {$b_6$};
    \node (b5) [below right=of b6] {$b_5$};
    \node (b4) [below right=of b5] {$b_4$};
    \node (b3) [below right=of b4] {$b_3$};
    \node (b2) [below left=of b3] {$b_2$};
    \node (b1) [below=of b2] {$b_1$};
    \node (top) [below right=of b1] {$\top$};
    \node (bot) [below left=of b1] {$\bot$};

    \draw[zeroarrow] (b6) -- (b5);
    \draw[zeroarrow] (b6) -- (b5);
    \draw[onearrow] (b5) -- (b4);
    \draw[onearrow] (b4) -- (b3);
    \draw[onearrow] (b3) -- (b2);
    \draw[zeroarrow] (b2) -- (b1);
    \draw[zeroarrow] (b1) -- (top);
    \draw[zeroarrow] (b3) -- (top);
    \draw[onearrow] (b2) -- (bot);
    \draw[onearrow] (b6) -- (bot);
    \draw[zeroarrow] (b5) -- (bot);
    \draw[zeroarrow] (b4) -- (bot);

    \end{tikzpicture}
}

\begin{example}[\cite{konstantinidis_applications_2013}]
We consider the language of BDDs over bit-vectors of length six. The following symbolic finite automaton accepts all strings that start by a bit-vector representing either of the numbers $6,14,22,38$ or $54$ followed by an arbitrary number of bit-vectors.

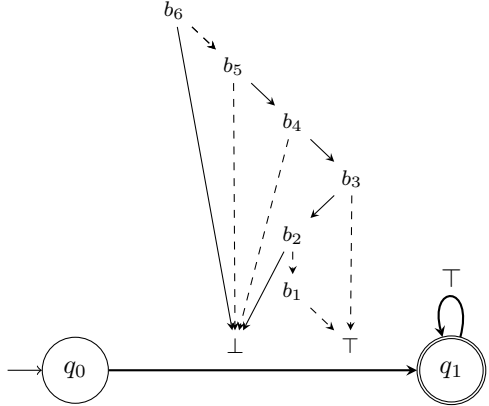
\centering
\begin{tikzpicture} [
    node distance = 5cm, 
    on grid, 
    auto,
    every loop/.style={stealth-}]
 
\node (q0) [state, 
    initial, 
    initial text = {}] {$q_0$};
 
\node (q1) [state,
    accepting,
    right = of q0] {$q_1$};
 
\path [-stealth, thick]
    (q0) edge node {\usebox{\mydiagram}}   (q1)
    (q1) edge [loop above]  node {$\top$}();
\end{tikzpicture}
\end{example}

\section{Feferman-Vaught Decomposition for SFAs}
\label{section:feferman}
Let $M = \left(\mathcal{A}, Q, q_{0}, F, \Delta\right)$ be a symbolic finite automaton and let $\psi_1, \ldots, \psi_k, \ldots$ be the atomic predicates in $\mathcal{A}$. The definition of symbolic finite automaton allows assuming that the set of these predicates is finite. 

\begin{lemma}
There exists a symbolic finite automaton $M' = \left(\mathcal{A}', Q, q_{0}, F, \Delta\right)$ such that $\mathcal{L}(M) = \mathcal{L}(M')$ and the cardinality of $\Psi_{\mathcal{A}'}$ is finite. 
\end{lemma}
\begin{proof}
The automaton has a finite number of transitions. We take $\Psi_{\mathcal{A}'}$ to be the Boolean closure of the predicates occurring in these transitions. It follows that $\Psi_{\mathcal{A}'}$ is a finite set. Otherwise, we define the components of $\mathcal{A}'$ as those in $\mathcal{A}$.  Since the automaton is unchanged, $\mathcal{L}(M) = \mathcal{L}(M')$. 
\end{proof}

Since $\Psi_{\mathcal{A}}$ can be assumed to be finite, it follows that the set of atomic predicates is finite too. In the remaining of the paper, we let $\phi_1,\ldots,\phi_k$ be the generators of the effective Boolean algebra used by the symbolic finite automaton $M$. Similarly, we let $\psi_1,\ldots,\psi_m$ denote the actual predicates used in the transitions of $M$.  We will decompose the study of $\mathcal{L}(M)$ into the study of the properties of the elements in $\mathcal{D}$ and the ordering properties induced by the transition structure of the automaton. Both kinds of properties will refer to sets of indices to stay in sync with each other \cite{wies_combining_2009}.

To specify the properties of the elements in $\mathcal{D}$, we use set interpretations of the form 
\begin{equation} \label{eq:1}
S = \set{ n \in \N | \psi(d(n)) } = \llbracket \psi \rrbracket
\end{equation}
where $d(n)$ is the $n$-th element occurring in $d \in \mathcal{D}^*$. These sets can be pictured via a Venn diagram of interpreted sets, such as the one in Figure~\ref{fig:venn}. Each formula in $\Psi_{\mathcal{A}}$ corresponds to a particular Venn region in this diagram and can be referred to using a Boolean algebra expression on the variables $S_1, \ldots, S_k$, thanks to the set interpretation (\ref{eq:1}).

\begin{figure} 
\centering
\begin{tikzpicture}[thick]
\draw (0,0) circle (1) node[above,shift={(0,1)}] {};
\draw (1.2,0) circle (1) node[above,shift={(0,1)}] {};
\draw (.6,-1.04) circle (1) node[shift={(1.1,-.6)}] {};

\node at (1.4,.2) {};
\node at (-.2,.2) {};
\node at (0.6,-1.4) {};

\draw [-to] (-.5,0) -- (.6,.4);
\draw [-to] (.6,.4) -- (1.5,.4);
\draw [-to] (1.5,.4) to[out=140,in=40,loop, distance=3cm] (1.5,.4);
\draw [-to] (1.5,.4) -- (.6,-.4);
\draw [-to] (.6,-.4) -- (2,-1.2);
\end{tikzpicture}
\caption{A table accepted by a symbolic automaton represented over an uninterpreted Venn diagram.}
\label{fig:run}
\end{figure}
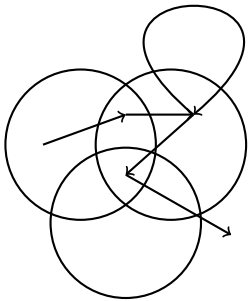

A concrete transition $q_1 \stackrel{d}{\to} q_2$ requires a value $d \in \mathcal{D}$. This value will lie in some elementary Venn region of the diagram in Figure~\ref{fig:venn}, i.e. in a set of the form $S_1^{\beta_1} \cap \ldots \cap S_k^{\beta_k}$ where $\beta = (\beta_1,\ldots,\beta_k) \in \{0,1\}^k, S^0 := S^c$ and $S^1 := S$. We will denote such Venn region with the bit-string $\beta$. To specify the transition structure of the automaton, what is relevant to us is the region of the Venn diagram, not the specific value that it takes there. It follows that a run of the automaton can be encoded as a sequence of bit-strings $\overline{t} = (t_1,\ldots,t_k) \in (\{0,1\}^{|\overline{t}|})^k$ and that these bit-strings only need to satisfy the propositional formulae corresponding to the predicates labelling the transitions of the automaton. Figure~\ref{fig:run} represents one such run over an uninterpreted Venn diagram.

\begin{example} \label{ex:prop}
If in Example~\ref{ex:smt-automaton} we take as atomic formulae the predicates $\psi_{\text{odd}}$ and $\psi_{> 0}$ then the formula $\psi_{\text{odd}} \land \psi_{> 0}(x)$, which labels the automaton transitions, corresponds to the propositional formula $S_1 \land S_2$. 
\end{example}

We denote by $L_1,\ldots,L_m$ such propositional formulae and by $M(L_1,\ldots,L_m)$ the set of bit-string runs accepted by $M$, which we call \textit{tables} \cite{kleene_representation_1956}.

\begin{lemma} \label{lem:automatonlanguage}
\begin{align*}
\mathcal{L}(M) = \Big\{ d \in \mathcal{D}^* \Big| &\exists \overline{t} \in M(L_1,\ldots,L_m).\\ &\bigwedge_{i = 1}^k S_i = \set{ n \in \N ; \phi_i(d(n)) } = \set{ n \in \N ; t_i(n) } \Big\}
\end{align*}
\end{lemma}
\begin{proof}
The proof uses the definition of $\mathcal{L}(M)$ and $M(L_1,\ldots,L_m)$. In one direction, one defines $\overline{t}$ from the membership of the values $d(i)$ in elementary Venn regions $\beta_i$. In the other direction, the definition of $M(L_1,\ldots,L_m)$ ensures that there is an accepting run corresponding to these values and any witness of the formula in the associated elementary Venn regions can be taken to conform the word $d$. 
\end{proof}

\begin{figure} 
\centering
\begin{tikzpicture}[thick]
\draw (0,0) circle (1) node[above,shift={(0,1)}] {};
\draw (1.2,0) circle (1) node[above,shift={(0,1)}] {};
\draw (.6,-1.04) circle (1) node[shift={(1.1,-.6)}] {};

\node at (-.2,.2) {$\phi_1$};
\node at (1.4,.2) {$\phi_2$};
\node at (0.6,-1.4) {$\phi_3$};
\end{tikzpicture}
\caption{A Venn diagram representing a finitely generated effective Boolean algebra with atomic predicates $\psi_1,\psi_2$ and $\psi_3$.}
\label{fig:venn}
\end{figure}
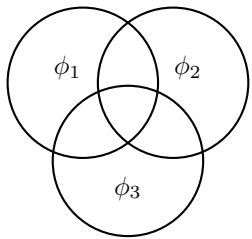

In the next sections, we make use of this decomposition to devise a decision procedure for symbolic finite automata, which, will refine the existing computational complexity results for the corresponding satisfiability problem.





\section{Decision Procedure for Satisfiability of SFAs}
\label{section:decision}
\begin{definition}
The satisfiability problem for a symbolic finite automaton $M$ is the problem of determining whether $\mathcal{L}(M) \neq \emptyset$.
\end{definition}

By Lemma~\ref{lem:automatonlanguage}, checking non-emptiness of the language of a symbolic finite automaton reduces to checking whether the following formula is true:
\begin{equation} \label{eq:table-power}
\begin{split}
\exists &S_1, \ldots, S_k. \exists d. \bigwedge_{i = 1}^k S_i = \set{ n \in \N ; \phi_i(d(n)) } \land \\ &\exists \overline{t} \in M(L_1,\ldots,L_m). \bigwedge_{i = 1}^k S_i = \set{ n \in \N ; t_i(n) }
\end{split}
\end{equation}

To establish the complexity of deciding formulae of the form~(\ref{eq:table-power}), we will have to analyse further the set $M(L_1,\ldots,L_m)$. Each table $\overline{t}$ in $M(L_1,\ldots,L_m)$ corresponds to a \textit{symbolic table} $\overline{s}$ whose entries are the propositional formulae that the bit-strings of $\overline{t}$ satisfy. More generally, these symbolic tables are generated by the symbolic automaton obtained by replacing the predicates of the symbolic automaton by propositional formulae. The set of symbolic tables accepted by the automaton $M$ is a regular set and will be denoted by $M_S(L_1,\ldots,L_m)$. 

\begin{example}
The automaton in Example~\ref{ex:smt-automaton} corresponds, according to Example~\ref{ex:prop}, to the symbolic automaton shown. 
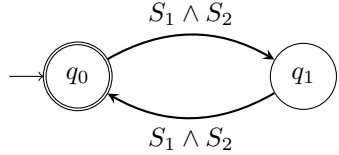
\begin{figure}[H]
\centering
\begin{tikzpicture} [
    node distance = 3cm, 
    on grid, 
    auto,
    every loop/.style={stealth-}]
 
\node (q0) [state, 
    initial, 
    accepting, 
    initial text = {}] {$q_0$};
 
\node (q1) [state,
    right = of q0] {$q_1$};
 
\path [-stealth, thick]
    (q0) edge[bend left] node {$S_1 \land S_2$}   (q1)
    (q1) edge[bend left] node {$S_1 \land S_2$}   (q0);
\end{tikzpicture}
\end{figure}
The symbolic tables generated by this automaton are of the form $((S_1 \land S_2)(S_1 \land S_2))^*$. The corresponding tables would be of the form $((1,1)(1,1))^*$.
\end{example}

Consider first the case where the propositional formulae $L_1,\ldots,L_m$ for the automaton $M$ denote disjoint Venn regions. In this case, all we need to do to check the satisfiability of formula~(\ref{eq:table-power}) is whether there exists a symbolic table $\overline{s}$ such that whenever the number of times a certain propositional letter occurs is non-zero, then the corresponding Venn region interpreted according to (\ref{eq:1}) has a satisfiable defining formula. From this, it follows that our decision procedure will need to compute the so-called Parikh image of the regular language $M_S(L_1,\ldots,L_m)$. 

\begin{definition}[Parikh Image] \label{def:parikh} ~\\
The Parikh image of $M_S(L_1,\ldots,L_m)$ is the set 
\[
\parikh(M_S(L_1,\ldots,L_m)) = \{ (|\overline{s}|_{L_1}, \ldots, |\overline{s}|_{L_m}) | \overline{s} \in M_S(L_1,\ldots,L_m) \} 
\]
where $|\overline{s}|_{L_i}$ denotes the number of occurrences of the propositional formula $L_i$ in the symbolic table $\overline{s}$. 
\end{definition}

We will use a description of the Parikh image in terms of linear-size existential Presburger arithmetic formulae first obtained by Seidl, Schwentick, Muscholl and Habermehl.

\begin{lemma}[\cite{seidl_counting_2004}] \label{lem:parikhlinear} 
The set $\parikh(M_S(L_1,\ldots,L_n))$ is definable by an existential Presburger formula $\rho$ of size $O(|M|)$ where $|M|$ is the number of symbols used to describe the automaton $M$.
\end{lemma}

When propositional letters denote overlapping Venn regions, a partitioning argument is required. This is formalised in Theorem~\ref{thm:tableelimination}. First, we fix some notation. We set $p_{\beta} := \bigcap_{i = 1}^k S_i^{\beta_i}$ where $\beta \in \{0,1\}^k$, $p_L := \bigcup\limits_{\beta \models L} p_{\beta}$ where $L$ is a propositional formula and $\varphi^{\beta}(d) := \bigwedge_{i = 1}^k  \varphi^{\beta(i)}_i(d)$. We write $S_1 \dot{\cup} S_2$ to denote 
the set $S_1 \cup S_2$ where it is known that $S_1 \cap S_2 = \emptyset$. Finally, we write $[n] := \set{1,\ldots,n}$.

\begin{theorem} \label{thm:tableelimination}
Formula~(\ref{eq:table-power}) is equivalent to the formula
\begin{equation} \label{eq:elimprop}
\begin{split}
\exists s \in& [m]. \sigma: [s] \hookrightarrow [m].\exists \beta_{1}, \ldots,\beta_{s} \in \{0,1\}^k. \bigwedge_{j = 1}^s \exists d. \phi^{\beta_{j}}(d) \land \\ \exists k_1,&\ldots,k_m. \exists S_1,\ldots, S_k, P_1,\ldots,P_s. \\ & 
  \rho(k_1,\ldots, k_m) \land \bigwedge_{i = 1}^s P_i \subseteq p_{L_{\sigma(i)}} \land  \cup_{i = 1}^m p_{L_i} = \dot{\cup}_{i = 1}^s P_i \land \\ 
  &\bigwedge_{i = 1}^s |P_i| = k_{\sigma(i)} \land \bigwedge_{i = 1}^s p_{\beta_i} \cap P_{i} \neq \emptyset
\end{split}
\end{equation}
where $\sigma$ is an injection from $\set{1,\ldots,s}$ to $\set{1,\ldots,m}$ and $\rho$ is the arithmetic expression in Lemma~\ref{lem:parikhlinear}.
\end{theorem}

Formula~(\ref{eq:elimprop}) has two parts. The first part corresponds to the subterm $\bigwedge_{j = 1}^s \exists d. \varphi^{\beta_{j}}(d)$ and falls within the theory of the elements in $\mathcal{D}$, $Th_{\exists^*}(\mathcal{D})$. The second part corresponds to the remaining subterm and falls within the quantifier-free first-order theory of Boolean Algebra with Presburger arithmetic ($\qfbapa$)\cite{kuncak_towards_2007}, which can be viewed as the monadic second order theory $Th_{\exists^*}^{mon}(\langle \mathbb{N}, \subseteq, \sim \rangle)$ where $\sim$ is the equicardinality relation between two sets. 

Formula~(\ref{eq:table-power}) is distilled from a non-deterministic decision procedure for the formulae of the shape (\ref{eq:table-power}). The existentially quantified variables $s, \sigma, \beta_{1}, \ldots,\beta_{s}$ are guessed by the procedure. These guessed values are then used by specialised procedures for $Th_{\exists^*}(\mathcal{D})$ and $Th_{\exists^*}^{mon}(\langle \mathbb{N}, \subseteq, \sim \rangle)$. For the convenience of the reader, we describe here what these values mean. The value of $s$ represents the number of Venn regions associated to the formulae $L_1,\ldots,L_m$ that will be non-empty. $\sigma$ indexes these non-empty regions. $\beta_1,\ldots,\beta_s$ are elementary Venn regions contained in the non-empty ones. 

The reason to introduce the partition variables $P_1,\ldots,P_s$ is that the Venn regions may overlap. 

\begin{example}
Consider the situation where $S_1 \land S_2$ and $S_2 \land S_3$ are two propositional formulae labelling the transitions of the symbolic automaton. These formulae correspond to the Venn regions $S_1 \cap S_2$ and $S_2 \cap S_3$, which share the region $S_1 \cap S_2 \cap S_3$. Given a model of $S_1, S_2$ and $S_3$, how do we guarantee that the indices in the region $S_1 \cap S_2 \cap S_3$ are consistent with a run of the automaton? For instance, the automaton may require one element in $S_1 \cap S_2$ and another in $S_2 \cap S_3$. Placing a single index in $S_1 \cap S_2 \cap S_3$ would satisfy the overall cardinality constraints, but not the fact that overall we need to have two elements. Trying to specify this in the general case would reduce to specifying an exponential number of cardinalities. 
\end{example}

We proceed next to the proof of the theorem. 

\begin{proof}[Proof of Theorem~\ref{thm:tableelimination}]
$\Rightarrow)$ If formula~(\ref{eq:table-power}) is satisfiable, then there are sets $S_1,\ldots,S_k$, a word $d$ and a table $\overline{t}$ satisfying
\begin{equation*} \label{eq:te1} 
\bigwedge_{i = 1}^k S_i = \Set{ n \in \N ; \phi_i(d) } \land \overline{t} \in M(L_1,\ldots,L_s) \land \bigwedge_{i = 1}^k S_i = \Set{ n \in \N ; t_i(n) }
\end{equation*}

Let $\overline{s} \in M(L_1,\ldots,L_s)$ be the symbolic table corresponding to $\overline{t}$. We define $k_i := |\overline{s}|_{L_i}, s = |\Set{ i | k_i \neq 0 }|$, $\sigma$ mapping the indices in $[s]$ to the indices of the terms for which $k_i$ is non-zero and $P_i = \set{ n \in \N ; \overline{s}(n) = L_{\sigma(i)} }$. It will be convenient to work out the following equalities:
\begin{align} \label{eq:te2}
\begin{split}
p_{L_i} &= \bigcup_{\beta \models L_i} \bigcap_{j = 1}^k S_j^{\beta_j} = \bigcup_{\beta \models L_i} \Set{ n \in \N | \bigwedge_{j = 1}^k t_j^{\beta_j}(n) } =  \Set{ n \in \N | \overline{t}(n) \models L_i } \\
p_{L_i} &= \bigcup_{\beta \models L_i} \bigcap_{j = 1}^k S_j^{\beta_j} = \bigcup_{\beta \models L_i} \Set{ n \in \N | \bigwedge_{j = 1}^k \phi_j^{\beta_j}(d) } =  \Set{ d \in \mathcal{D} | L_i(\overline{\phi}(d)) }
\end{split}
\end{align}
where $L_i(\overline{\phi}(d(n)))$ is the propositional formula obtained by substituting set variables by the formulae $\phi_i(d(n))$. We now deduce formula~(\ref{eq:elimprop}):
\begin{itemize}
    \renewcommand\labelitemi{-}
    \item $\rho(k_1,\ldots,k_m)$: from $\overline{s} \in P(L_1,\ldots,L_m)$, we have that  
    \[
    (k_1,\ldots,k_m) \in \parikh(M_S(L_1,\ldots,L_m))
    \]
    and therefore $\rho(k_1,\ldots,k_m)$.
    \item $P_i \subseteq p_{L_{\sigma(i)}}$: since $\overline{s}$ corresponds to $\overline{t}$, for all $n \in \N$ we have $\overline{t}(n) \models \overline{s}(n)$ and the inclusion follows from the definition of $P_i$ and equation~\ref{eq:te2}.
    \item $|P_i| = k_{\sigma(i)}$: since $|P_i| = \Big|\Set{ n \in \N | \overline{s}(n) = L_{\sigma(i)} }\Big| = |\overline{s}|_{L_{\sigma(i)}} = k_{\sigma(i)}$.
    \item Each pair of sets $P_i, P_j$ with $i < j$ is disjoint: 
    \begin{align*}
    P_i \cap P_j &= \Set{ n \in \N | \overline{s}(n) = L_{\sigma(i)} } \cap \Set{ n \in \N | \overline{s}(n) = L_{\sigma(j)} } = \\ &= \Set{ n \in \N | \overline{s}(n) = L_{\sigma(i)} = L_{\sigma(j)} } = \emptyset
    \end{align*}
    using that the letters $L$ are chosen to be distinct and that $\sigma$ is an injection (so $\sigma(i) \neq \sigma(j)$). 

    \item $p_{L_1} \cup \ldots \cup p_{L_m} = P_1 \dot{\cup} \ldots \dot{\cup} P_s$: since by definition $P_i = \Set{ n \in \N | \overline{s}(n) = L_{\sigma(i)} }$, $p_{L_i} = \Set{ n \in \N | \overline{t}(n) \models L_i }$ and by definition of $\sigma$ it follows that the only letters that can appear in $\overline{s}$ are $L_{\sigma(1)}, \ldots, L_{\sigma(s)}$. Thus, we have $p_{L_1} \cup \ldots \cup p_{L_m} = [1,|\overline{t}|] = [1,|\overline{s}|] = P_1 \dot{\cup} \ldots \dot{\cup} P_s$. 
    
    \item There exists $\beta_1,\ldots,\beta_s \in \{0,1\}^k$, such that $\bigwedge_{i = 1}^s P_{\beta_i} \cap P_{i} \neq \emptyset$: note that $P_i \neq \emptyset$ by definition of $\sigma$. Thus, there must exist some $\beta_i$ such that $p_{\beta_i} \cap P_{i} \neq \emptyset$. We pick any such $\beta_i$.
    \item $\bigwedge_{j = 1}^s \exists d.  \varphi^{\beta_j}(d)$: follows from $p_{\beta_{j}} \cap P_j \neq \emptyset$ and formula~(\ref{eq:te2}).
\end{itemize}

$\Leftarrow)$ Conversely, if formula~(\ref{eq:elimprop}) is satisfiable, then there is an integer $s \in [n]$, an injection $\sigma: [s] \hookrightarrow [n]$, bit-strings $\beta_{1},\ldots,\beta_{s} \in \{0,1\}^k$, integers $k_1,\ldots,k_m$ and sets $S_1,\ldots,S_k,P_1,\ldots,P_s$ satisfying  
\begin{equation} 
\begin{split}
&\bigwedge_{j = 1}^s \exists d.   \varphi^{\beta_{j}}(d) \land 
  \rho(k_1,\ldots, k_m) \land \bigwedge_{i = 1}^s P_i \subseteq p_{L_{\sigma(i)}} \land  \cup_{i = 1}^m p_{L_i} = \dot{\cup}_{i = 1}^s P_i \land \\
 &\bigwedge_{i = 1}^s |P_i| = k_{\sigma(i)} \land \bigwedge_{i = 1}^s p_{\beta_{i}} \cap P_{i} \neq \emptyset 
\end{split}
\end{equation}
From $
\psi(k_1,\ldots, k_n)
$ follows that there is a symbolic table $\overline{s} \in M_S(L_1,\ldots,L_m)$ such that $|\overline{s}|_{L_i} = k_i$ for each $L_i \in \set{L_1,\ldots,L_m}$. From formula~(\ref{eq:te2}) and
\begin{equation*}
p_{L_1} \cup \ldots \cup p_{L_m} = P_1 \dot{\cup} \ldots \dot{\cup} P_s \land 
\bigwedge_{i = 1}^s P_{i} \subseteq p_{L_{\sigma(i)}} \land \bigwedge_{i = 1}^s |P_i| = k_{\sigma(i)} 
\end{equation*}
follows that we can replace the formulae $L_i$ occurring in the symbolic table $\overline{s}$ by the bit-strings representing the elementary Venn regions to which the indices of the sets $P_i$ belong. Moreover, thanks to the condition $\bigwedge_{i = 1}^s p_{\beta_{i}} \cap P_{i} \neq \emptyset$ follows that we can replace the letters $L_i$ by the bit-strings $\beta_i$, defining $\overline{t}$ as $\overline{t}(n) = 
\begin{cases}
\beta_{i} & \text{if } n \in P_{i}
\end{cases}$. In this way, we obtain a table $\overline{t} \in M(L_1,\ldots,L_s)$. We then define the corresponding word over $\mathcal{D}$, thanks to the property $\bigwedge_{i = 1}^s \exists d. \phi^{\beta_{i}}(d)$. Naming the witnesses of these formulae as $d_{i}$, we define $d(n) = 
\begin{cases}
d_{i} & \text{if } n \in P_{i}
\end{cases}$. To conclude, note that:
\[
\Set{ n \in \N | t_j(n) } = \cup_{\Set{ 1 \le i \le k | \beta_{i}(j) = 1}} P_{i} = \Set{ n \in \N | \phi_j(d(n)) }
\]
Thus, we have that formula~(\ref{eq:table-power}) is satisfied by the set variables 
\[
S_j := \Set{ n \in \N | t_j(n) } = \Set{ n \in \N | \phi_j(d(n)) }
\]
\end{proof}

\section{Quantifier-free Boolean Algebra with Presburger Arithmetic}
\label{section:with}
The arguments following the statement of Theorem~\ref{thm:tableelimination} sketch a non-deterministic procedure for the satisfiability problem of symbolic finite automata, based on the existence of decision procedures for $Th_{\exists^*}(\mathcal{D})$ and $Th_{\exists^*}^{mon}(\langle \mathbb{N}, \subseteq, \sim \rangle)$. In this section, we recall the non-deterministic polynomial time decision procedure for $Th_{\exists^*}^{mon}(\langle \mathbb{N}, \subseteq, \sim \rangle)$. As a consequence, we obtain Corollary~\ref{cor:complexity} which situates the decision problem of symbolic finite automata in the classical complexity hierarchy. This section should also prepare the reader for the extension of these results, where the automaton can require linear arithmetic constraints on the cardinalities of the effective Boolean algebra. This extension is carried out in Section~\ref{section:cardinalities}. 

Instead of working with $Th_{\exists^*}^{mon}(\langle \mathbb{N}, \subseteq, \sim \rangle)$ directly, we use the logic $\qfbapa$ \cite{kuncak_towards_2007} which has the same expressive power \cite[Section~2]{kuncak_deciding_2006}. The syntax of $\qfbapa$ is given in Figure~\ref{fig:qfbapa-syntax}. The meaning of the syntax is as follows. $F$ presents the Boolean structure of the formula, $A$ stands for the top-level constraints, $B$ gives the Boolean restrictions and $T$ the Presburger arithmetic terms. The operator $\text{dvd}$ stands for the divisibility relation and $\mathcal{U}$ represents the universal set. The remaining interpretations are standard.

\begin{figure}[!ht]
\centering
\begin{align*}
F & ::= A \, | \, F_1 \land F_2 \, | \, F_1 \lor F_2 \, | \, \lnot F \\
A & ::= B_1 = B_2 \, | \, B_1 \subseteq B_2 \, | \, T_1 = T_2 \, | \, T_1 \le T_2 \, | \, K \text{ dvd } T \\
B & ::= x \, | \, \emptyset \, | \, \mathcal{U} \, | \, B_1 \cup B_2 \, | \, B_1 \cap B_2 \, | \, B^c \\
T & ::= k \, | \, K \, | \,  T_1 + T_2 \, | \, K \cdot T \, |  \, |B| \\
K & ::= \ldots \, | \, -2 \, | \, -1 \, | \, 0 \, | \, 1 \, | \, 2 \, | \, \ldots
\end{align*}
\caption{$\qfbapa$'s syntax}
\label{fig:qfbapa-syntax}
\end{figure}

The satisfiability problem of this logic is reducible to propositional satisfiability in polynomial time. Our proofs will rely on the method of \cite{kuncak_towards_2007}, which we sketch briefly here. The basic argument to establish a $\np$ complexity bound on the satisfiability problem of $\qfbapa$ is based on a theorem by Eisenbrand and Shmonin \cite{eisenbrand_caratheodory_2006}, which in our context says that any element of an integer cone can be expressed in terms of a polynomial number of generators. Figure~\ref{fig:pa-verifier} gives a verifier for this basic version of the algorithm. The algorithm uses an auxiliary verifier $V_{PA}$ for the quantifier-free fragment of Presburger arithmetic. The key step is showing equisatisfiability between 2.(b) and 2.(c). If $x_1, \ldots, x_k$ are the variables occurring in $b_0, \ldots, b_p$ then we write $p_\beta = \bigcap\limits_{i = 1}^k x_i^{e_i}$ for $\beta = (e_1,\ldots,e_k) \in \{0,1\}^k$ where we define $x^1 := x$ and $x^0 := \mathcal{U} \setminus x$ as before. If we define $\llbracket b_i \rrbracket_{\beta_j}$ as the evaluation of $b_i$ as a propositional formula with the assignment given in $\beta$ and introduce variables $l_\beta = |p_\beta|$, then $|b_i| =  \sum\limits_{j = 0}^{2^e-1} \llbracket b_i \rrbracket_{\beta_j} l_{\beta_j}$, so the restriction $\bigwedge\limits_{i = 0}^p |b_i| = k_i$ in 2.(b) becomes $\bigwedge\limits_{i = 0}^p \sum\limits_{j = 0}^{2^e-1} \llbracket b_i \rrbracket_{\beta_j} l_{\beta_j} = k_i$ which can be seen as a linear combination in the set of vectors $
\{(\llbracket b_0 \rrbracket_{\beta_j}, \ldots, \llbracket b_p \rrbracket_{\beta_j}). j \in \{0, \ldots, 2^e-1\}\} 
$. Eisenbrand-Shmonin's result allows then to derive 2.(c) for $N$ polynomial in $|x|$. In the other direction, it is sufficient to set $l_{\beta_j} = 0$ for $j \in \{0, \ldots, 2^e-1\} \setminus \{i_1, \ldots, i_N\}$. Thus, we have:

\begin{figure*}[ht!]
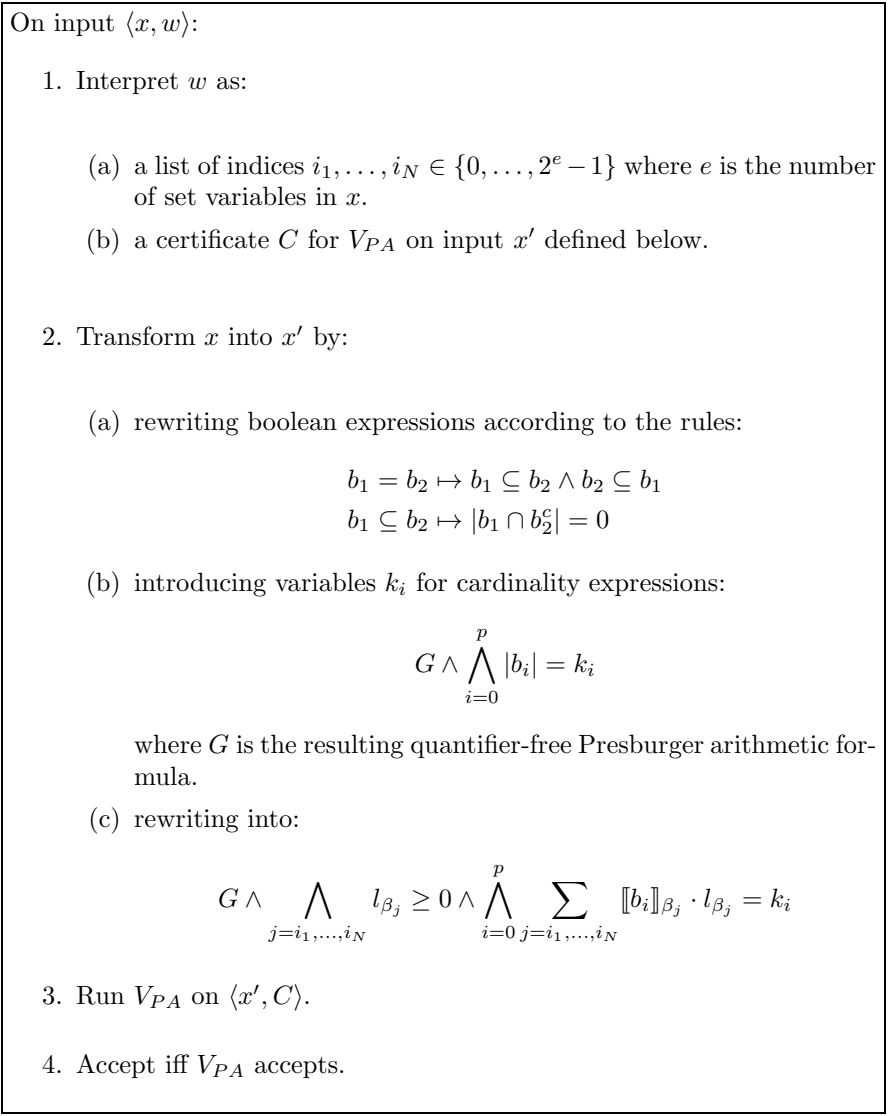

\fbox{\parbox{.95\textwidth}{
On input $\langle x, w \rangle$:

\begin{enumerate}
\setlength\itemsep{1em}
\item Interpret $w$ as:

\vspace{1em}

\begin{enumerate}
    \item a list of indices $i_1, \ldots, i_N \in \{0, \ldots, 2^e-1 \}$ where $e$ is the number of set variables in $x$.
    \item a certificate $C$ for $V_{PA}$ on input $x'$ defined below.
\end{enumerate}

\vspace{.5em}

\item Transform $x$ into $x'$ by:

\vspace{1em}

\begin{enumerate}
    \item rewriting boolean expressions according to the rules: \begin{align*}
    b_1 = b_2 & \mapsto b_1 \subseteq b_2 \land b_2 \subseteq b_1 \\
    b_1 \subseteq b_2 & \mapsto |b_1 \cap b_2^c| = 0
    \end{align*}
    
    \item introducing variables $k_i$ for cardinality expressions: $$G \land \bigwedge_{i = 0}^{p} |b_i| = k_i$$ where $G$ is the resulting quantifier-free Presburger arithmetic formula.
    
    \item rewriting into:
    $$ G \land \bigwedge\limits_{j = i_1, \ldots, i_N} l_{\beta_j} \ge 0 \land \bigwedge_{i = 0}^{p} \sum_{j = i_1, \ldots, i_N} \llbracket b_i \rrbracket_{\beta_j} \cdot l_{\beta_j} = k_i$$
\end{enumerate}

\item Run $V_{PA}$ on $\langle x', C \rangle$.

\item Accept iff $V_{PA}$ accepts.
\end{enumerate}
}}
\caption{Verifier for $\qfbapa$}
\label{fig:pa-verifier}
\end{figure*}

\begin{theorem}[{\cite{kuncak_towards_2007}}] \label{thm:qfbapa-complexity}
The satisfiability problem of $\qfbapa$ is in NP. 
\end{theorem}

From Theorems~\ref{thm:tableelimination} and \ref{thm:qfbapa-complexity}, we obtain the following improvement of \cite[Theorem~2.8]{dantoni_automata_2021}:

\begin{corollary} \label{cor:complexity}
Let $Th_{\exists^*}(\mathcal{D})$ be the existential first-order theory of the formulae used in the transitions of the symbolic finite automaton $M$. 

\begin{itemize}
\item If $Th_{\exists^*}(\mathcal{D}) \in \p$ then $\mathcal{L}(M) \neq \emptyset \in \np$. 

\item If $Th_{\exists^*}(\mathcal{D}) \in \cc$ for some $\cc \supseteq \np$ then $\mathcal{L}(M) \neq \emptyset \in \cc$. 
\end{itemize}
\end{corollary}

\section{Decision Procedure for Satisfiability of SFAs with Cardinalities}
\label{section:cardinalities}
We now consider the following generalisation of the language of a finite symbolic automaton from Lemma~\ref{lem:automatonlanguage}.

\begin{definition}
A symbolic finite automaton  with cardinalities accepts a language  of the form:
\[
\mathcal{L}(M) = \left 
\{ d \in \mathcal{D}^* \middle \vert \begin{array}{l}
F(S_1,\ldots,S_k) \land \bigwedge_{i = 1}^k S_i = \Set{ n \in \N | \phi_i(d(n)) } \land \\ \exists \overline{t} \in P(L_1,\ldots,L_m). \bigwedge_{i = 1}^k S_i = \Set{ n \in \N | t_i(n) } 
\end{array}
\right\}
\]
where $F$ is a formula from $\qfbapa$. 

Thus, checking non-emptiness of the language of a symbolic finite automaton with cardinalities reduces to checking whether the following formula is true:
\begin{equation} \label{eq:table-power-card}
\begin{split}
\exists S_1, \ldots, S_k. &F(S_1,\ldots,S_k) \land \\ &\exists d. \bigwedge_{i = 1}^k S_i = \Set{ n \in \N | \phi_i(d(n)) } \land \\ &\exists \overline{t} \in M(L_1,\ldots,L_m) \land \bigwedge_{i = 1}^k S_i = \Set{ n \in \N | t_i(n) }
\end{split}
\end{equation}
\end{definition}

To show that Theorem~\ref{thm:tableelimination} and Corollary~\ref{cor:complexity} stay true with linear arithmetic constraints on the cardinalities, we need to repeat part of the argument in Theorem~\ref{thm:tableelimination} since if $F$ denotes the newly introduced $\qfbapa$ formula and $G,H$ are the formulae shown equivalent in Theorem~\ref{thm:tableelimination}, then from: 
\[
\exists S_1,\ldots,S_k. F(S_1,\ldots,S_k) \land G(S_1,\ldots,S_k)
\]
and
\[
\Big[ \exists S_1,\ldots,S_k. G(S_1,\ldots,S_k) \Big] \iff \Big[ \exists S_1,\ldots,S_k. H(S_1,\ldots,S_k) \Big]
\]
it does not follow that $\exists S_1,\ldots,S_k. F(S_1,\ldots,S_k) \land H(S_1,\ldots,S_k)$. Instead, the algorithm derives the cardinality constraints from each theory and then uses the sparsity of solutions \textit{over the satisfiable regions}. In the proof, we set $\llbracket \beta_j \models b_i \rrbracket$ to be one, if the bit-string $\beta_j$ satisfies the Boolean expression $b_i$ as a propositional assignment and zero otherwise. We also write $l_{\beta} = |p_{\beta}|$ for $\beta \in \{0,1\}^k$. 

\begin{theorem}
Formula~(\ref{eq:table-power-card}) is equivalent to:
\begin{equation} \label{eq:elimpropcard}
\begin{split}
\exists N \le p(|F|), &\exists s \in [m]. \sigma: [s] \hookrightarrow [m]. 
\exists \beta_{1}, \ldots,\beta_{N} \in \{0,1\}^k. \bigwedge_{j = 1}^N \exists d. \phi^{\beta_{j}}(d) \land \\ \exists k_1,\ldots,k_m.& \exists S_1,\ldots, S_k, P_1,\ldots,P_s. \\ 
  & \rho(k_1,\ldots, k_m) \land \bigwedge_{i = 1}^s P_i \subseteq p_{L_{\sigma(i)}} \land  \cup_{i = 1}^m p_{L_i} = \dot{\cup}_{i = 1}^s P_i \land \\ 
  &\bigwedge_{i = 1}^s |P_i| = k_{\sigma(i)} \land \cup_{i = 1}^N p_{\beta_i} = \dot{\cup}_{i = 1}^s P_{i}
\end{split}
\end{equation}
where $p$ is a polynomial and $|F|$ is the number of symbols used to write $F$. 
\end{theorem}
\begin{proof}
$\Rightarrow)$ If formula~(\ref{eq:table-power-card}) is true, then there are sets $S_1,\ldots,S_k$, a finite word $d$ and a table $\overline{t}$ such that:
\begin{align} 
\label{eq:te2card}
\begin{split}
&F(S_1,\ldots,S_k) \land \bigwedge_{i = 1}^k S_i = \Set{ n \in \N | \phi_i(d(n)) } \land \\ 
&\overline{t} \in M(L_1,\ldots,L_m) \land \bigwedge_{i = 1}^k S_i = \Set{ n \in \N | t_i(n) }
\end{split}
\end{align}
Thus, there exists a symbolic table $\overline{s} \in M_S(L_1,\ldots,L_s)$ corresponding to $\overline{t}$. We define $k_i := |\overline{s}|_{L_i}, s = |\Set{ i | k_i \neq 0 }|$, $\sigma$ maps the indices in $[s]$ to the indices of the terms for which $k_i$ is non-zero and $P_i = \Set{ n \in \N | \overline{s}(n) = L_{\sigma(i)} }$. As in Theorem~\ref{thm:tableelimination}, we have the equalities $p_{L_i} = \Set{ n \in \N | \overline{t}(n) \models L_i }$, $p_{L_i} = \Set{ n \in \N | L_i(\overline{\phi}(d)) }$ and we can show that the following formula holds:
\begin{equation} \label{eq:derivedcond}
\begin{split}
&\rho(k_1,\ldots, k_m) \land \bigwedge_{i = 1}^m P_i \subseteq p_{L_{\sigma(i)}} \land \cup_{i = 1}^m p_{L_i} = \dot{\cup}_{i = 1}^s P_i \land \\ & \bigwedge_{i = 1}^s |P_i| = k_{\sigma(i)} \land F(S_1,\ldots,S_k)
\end{split}
\end{equation}
We need to find a sparse model of (\ref{eq:derivedcond}). To achieve this, we follow the methodology in Theorem~\ref{thm:qfbapa-complexity}. This leads to a system of equations of the form:
\[
\exists c_1,\ldots,c_p. G \land \sum_{j = 0}^{2^e - 1} 
\begin{pmatrix}
\llbracket \beta_j \models b_0 \rrbracket \\
\cdots \\
\llbracket \beta_j \models b_p \rrbracket
\end{pmatrix}\cdot l_{\beta_j} = 
\begin{pmatrix}
c_1 \\
\ldots \\
c_p
\end{pmatrix}
\]
We remove those elementary Venn regions where $l_{\beta} = 0$. This includes regions whose associated formula in the interpreted Boolean algebra is unsatisfiable, and regions corresponding to table entries not occurring in $\overline{t}$. This transformation gives a reduced set of indices $\mathcal{R}$ participating in the sum. 

Using Eisenbrand-Shmonin's theorem, we have a polynomial (in the size of the original formula) family of Venn regions $\beta_1, \ldots,\beta_N$ and corresponding cardinalities $l_{\beta_1}',\ldots,l_{\beta_N}'$, which we can assume to be non-zero, such that
\begin{equation} \label{eq:reduced}
\exists c_1,\ldots,c_p. G \land \sum_{\beta \in \set{\beta_1, \ldots,\beta_N} \subseteq \mathcal{R}}
\begin{pmatrix}
\llbracket \beta_j \models b_0 \rrbracket \\
\cdots \\
\llbracket \beta_j \models b_p \rrbracket
\end{pmatrix}\cdot l_{\beta_j}' = 
\begin{pmatrix}
c_1 \\
\ldots \\
c_p
\end{pmatrix}
\end{equation}


The satisfiability of formula~(\ref{eq:reduced}) implies the existence of sets of indices $p_{\beta}'$ satisfying the conditions derived in formula~(\ref{eq:derivedcond}). However, it does not imply which explicit indices belong to these sets and which are the contents corresponding to each index. From the condition 
\begin{equation*}
\psi(k_1,\ldots, k_n) \land \bigwedge_{i = 1}^n P_i' \subseteq p_{L_{\sigma(i)}}' \land \cup_{i = 1}^n p_{L_i}' = \dot{\cup}_{i = 1}^s P_i' \land \bigwedge_{i = 1}^s |P_i'| = k_{\sigma(i)} 
\end{equation*}
follows that there is a symbolic table $\overline{s}'$ satisfying $M_S(L_1,\ldots,L_n)$ with $k_{\sigma(i)}$ letters $L_{\sigma(i)}$ and that these letters are made concrete by entries in $P_i'$ for each $i \in \{1,\ldots,s\}$. We take the Venn regions $\beta \in \{\beta_1,\ldots,\beta_N\}$ such that $P_i' \supseteq p_{\beta}$ and label the corresponding entries in $\overline{s}'$ with $\beta$. In this way, we obtain a corresponding concrete table $\overline{t}'$. This makes the indices in each Venn region concrete. To make the contents of the indices concrete, note that for each $\beta \in \mathcal{R}$, since $l_\beta \neq 0$, the formula $\exists d.
\phi^{\beta}(d)$ is true. In particular, this applies to each $\beta \in \{\beta_1, \ldots, \beta_N\}$. Thus, we obtain witnesses $d_1,\ldots,d_N$. We form a word by replacing each letter $\beta$ in $\overline{t}'$ by the corresponding value $d_{\beta}$.

\vspace{.5em}

$\Leftarrow)$ If formula~(\ref{eq:elimpropcard}) is true, then there is $N \le p(|F|)$ where $p$ is a polynomial, $s \in [m]$, $\beta_1,\ldots,\beta_N \in \{0,1\}^k, k_1,\ldots,k_m \in \N$ and sets $S_1,\ldots,S_k, P_1,\ldots, P_s$ such that
\begin{align*}
\bigwedge_{j = 1}^N \exists d. \phi^{\beta_{j}}(d) \land &\rho(k_1,\ldots, k_m) \land \bigwedge_{i = 1}^s P_i \subseteq p_{L_{\sigma(i)}} \land  \cup_{i = 1}^m p_{L_i} = \dot{\cup}_{i = 1}^s P_i \land \\ 
  &\bigwedge_{i = 1}^s |P_i| = k_{\sigma(i)} \land \cup_{i = 1}^N p_{\beta_i} = \dot{\cup}_{i = 1}^s P_{i}
\end{align*}

From $
\rho(k_1,\ldots, k_n)
$ follows that there is a symbolic table $\overline{s} \in R(L_1,\ldots,L_m)$ such that $|\overline{s}|_{L_i} = k_i$ for each $L_i \in \set{L_1,\ldots,L_m}$. From formula~\ref{eq:te2card} and
\begin{equation*}
p_{L_1} \cup \ldots \cup p_{L_m} = P_1 \dot{\cup} \ldots \dot{\cup} P_s \land 
\bigwedge_{i = 1}^s P_{i} \subseteq p_{L_{\sigma(i)}} \land \bigwedge_{i = 1}^s |P_i| = k_{\sigma(i)} 
\end{equation*}
follows that we can replace the formulae $L_i$ occurring in the symbolic table $\overline{s}$ by the bit-strings representing the elementary Venn regions to which the indices of the sets $P_i$ belong. Moreover, thanks to the condition $\cup_{i = 1}^N p_{\beta_i} = \dot{\cup}_{i = 1}^s P_{i}$, it follows that we can replace the letters $L_i$ by the bit-strings $\beta_i$. In this way, we obtain a table $\overline{t} \in R(L_1,\ldots,L_m)$. We then define the corresponding word over $\mathcal{D}$, thanks to the property $
\bigwedge_{i = 1}^N \exists d. \phi^{\beta_{i}}(d) 
$. To conclude, note that:
\[
\Set{ n \in \N | t_j(n) } = \cup_{\Set{ i | \beta_{i}(j) = 1}} P_{i} = \Set{ n \in \N | \phi_j(d(n)) }
\]
Thus, we have that formula~\ref{eq:table-power} is satisfied by the set variables 
\[
S_j := \Set{ n \in \N | t_j(n) } = \Set{ n \in \N | \phi_j(d(n)) }
\]
\end{proof}

We can thus formulate the analogous to Corollary~\ref{cor:complexity} in the case of finite symbolic automata with cardinalities.

\begin{corollary} \label{cor:complexitycard}
Let $Th_{\exists^*}(\mathcal{D})$ be the theory of the formulae used in the transitions of a symbolic finite automaton with cardinality constraints.

\begin{itemize}
\item If $Th_{\exists^*}(\mathcal{D}) \in \p$ then $\mathcal{L}(M) \neq \emptyset \in \np$. 

\item If $Th_{\exists^*}(\mathcal{D}) \in \cc$ for some $\cc \supseteq \np$ then $\mathcal{L}(M) \neq \emptyset \in \cc$. 
\end{itemize}
\end{corollary}

\section{Conclusion}
\label{section:conclusion}
We have revisited the model of symbolic finite automata as it was reintroduced in \cite{veanes_rex_2010}. We have obtained tight complexity bounds on their satisfiability problem. Our methodology follows the Feferman-Vaught decomposition technique in that it reduces the satisfiability problem of the automaton to the satisfiability problem of the existential first-order theory of the characters accepted by the automaton and the satisfiability problem of the existential monadic second-order theory of the indices. 

To combine these two distinct theories we use the ideas from the combination method through sets and cardinalities of Wies, Piskac and Kun\v{c}ak \cite{wies_combining_2009} and the computation of an equivalent linear-sized existentially quantified Presburger arithmetic formula from the Parikh image of a regular language by Seidl, Schwentick, Muscholl and Habermehl \cite{seidl_counting_2004}. A crucial step in the proofs is a partitioning argument for the underlying Venn regions. We profit from the analysis in \cite{kuncak_towards_2007} to extend our arguments to the satisfiability problem of finite symbolic automata that consider linear arithmetic restrictions over the cardinalities of the Boolean algebra associated with the symbolic finite automaton. 

In future work, we plan to extend our methods to other variants of symbolic automata to which we believe similar techniques may be applicable. Another interesting research direction would be to consider extensions of the language that allow free variables in set interpretations of the form~(\ref{eq:1}), which seems to have applications to various satisfiability problems.

\end{document}